A. Yazdani * and L. Shojai

# Solution of a Scalar Convection-Diffusion Equation Using FEMLAB

**Abstract** A steady scalar convection-diffusion problem has been studied for one and two dimensional cases. The major problem of unrealistic oscillations of the convection dominated problems is relaxed thanks to the wide range of the elements FEMLAB 3.1 benefits. The FEMLAB 3.1 solution has been presented for the problems, unique features and illustrations of the software have been used and results have been tested against analytic solution.

## 1. Introduction

Processes involving a combination of convection and diffusion are ubiquitously found in physical and engineering problems. These problems occur in many applications such as in the transport of air and ground water pollutants, oil reservoir flow, in the modelling of semiconductors, and so forth. Convection is a physical process by which some property is transported by the ordered motion of the flow, while diffusion is the physical process by which the property is transported by the random motion of the molecules of the fluid. The behaviour of fluid undergoing mass, vorticity, or forced heat transfer is described by a set of partial differential equations which are mathematical formulations of one or more of the conservation laws of physics. These laws include those of conservation of mass, momentum, and energy. The numerical solution of a convection diffusion equations whose first derivative have large coefficients (convection dominated) presents difficulties such as parasitic oscillation and instability. Several finite element treatments of the problem have ever been tried and developed, including upwinding techniques, Petrov-Galerkin approach and artificial diffusivity method, and the more recent stabilized methods. FEMLAB 3.1, takes the advantage of employing these methods in a very straightforward and user-friendly way. The obtained results are therefore, as reliable and accurate as the results based on the most recent techniques, bearing in mind that all code developments and programming works are already done.

## 2. Statement of the Problem

The general steady linear problem on a bounded domain is of the form

$$-\varepsilon \nabla \cdot (a\nabla u) + \nabla \cdot (\vec{b}u) + cu = S \; in \; \Omega \qquad (1\text{-}2)$$

with boundary conditions

$$u = u_B \; on \; \partial\Omega_D, \; \frac{\partial u}{\partial n} = 0 \; on \; \partial\Omega_N \qquad (2\text{-}2)$$

where $\partial\Omega_D, \partial\Omega_N$ form a partition of the boundary of $\Omega$ in which $\partial\Omega_D$ is non-empty. We usually assume that the advective velocity field $\vec{b}$ is incompressible, $\nabla \vec{b} = 0$, so that the convective term can also be written $\vec{b}.\nabla u$, and also $a(x) \geq 1, c(x) \geq 0$ while $\varepsilon$ is a small positive constant [2].

First we consider the one dimensional differential equation:

$$\frac{d^2u}{dx^2} - \vec{b}(x)\frac{du}{dx} = S(x) \; in \; \Omega = [0,1] \qquad (3\text{-}2)$$

subject to the given boundary conditions $u(0) = 1, \; u(1) = 0$. We assume that the convective term $\vec{b}(x)$ is a constant and there is no source term, i.e. $S(x) = 0$. In this case the analytic solution of equation (3-2) is given as:

$$u(x) = \frac{e^{bx} - e^b}{1 - e^b}. \qquad (4\text{-}2)$$

Numerical schemes successfully cope with such a simplified linearized equation.

Setting b=50, and using predefined cubic Lagrange element, FEMLAB 3.1 returns: number of elements: 30, number of degrees of freedom: 91, solution time: 0.032 Seconds, for which the stationary analysis, Coefficient forms, PDE module, has been used. Figure 1. depicts similarity of analytic solution and FEMLAB result.

A. Yazdani *  L. Shojai
Department of Chemical Engineering,
Loughborough University,
Loughborough, LE11 3TU, UK
A.Yazdani@lboro.ac.uk

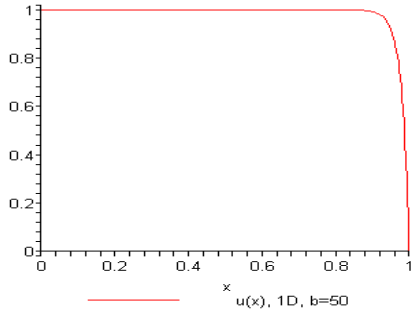

1.a. Analytic result b=50

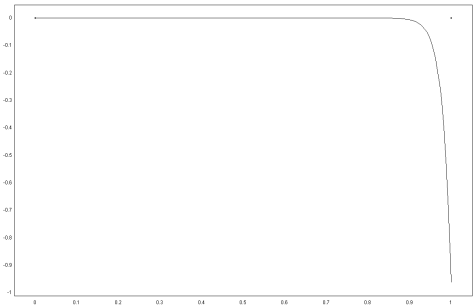

1.b. FEMLAB result b=50

**Figure 1. Comparison of analytic and FEMLAB results**

The solution (using cubic Lagrange elements) is fast, accurate and highly reliable compared to the analytic solution.

## 3. Two dimensional test problem

Two dimensional convection-diffusion problem is represented by:

$$\frac{\partial^2 u}{\partial x^2}+\frac{\partial^2 u}{\partial y^2}-b_1\frac{\partial u}{\partial x}-b_2\frac{\partial u}{\partial y}=S. \qquad (3\text{-}1)$$

Unlike the one dimensional case, it is not very easy to invent a range of two dimensional problems with ready analytical solutions. As a test problem we study the convection-diffusion model of (3-1) with constant $b$ and no source term, over the unit square. For this purpose we set $b_1=b_2$, $S=0$, $\Omega=[0,1]\times[0,1]$ subject to the following Dirichlet boundary conditions:

$$\begin{cases} u(x,1)=0=u(1,y)\\ u(x,0)=1=u(0,y)\\ u(0,1)=0.5=u(1,0)\end{cases} \qquad (3\text{-}2)$$

We shall simply write:

$$\begin{cases}\nabla^2 u - b\nabla u = 0\\ B.C.\end{cases} \qquad (3\text{-}3)$$

(which can be interpreted as the energy conservation equation with no heat source term). By the method of separation of variables the analytical solution of (3-2) is given by:

$$u(x,y)=\sum_{n=1}^{\infty}\left|\frac{(1-(-1)^n e^{\frac{-b}{2}})8n\pi e^{\frac{b(x+y)}{2}}}{\sinh(\frac{\sqrt{A_n}}{2})(b^2+4n^2\pi^2)}\right.$$
$$\left|\sin(n\pi x)\sinh(\frac{\sqrt{A_n}(1-y)}{2})\right. \qquad (3\text{-}4)$$
$$\left.+\sin(n\pi y)\sinh(\frac{\sqrt{A_n}(1-x)}{2})\right|$$

where $A_n = 2b^2 + 4n^2\pi^2 > 0$. The common higher degree interpolation functions give acceptable solution for value of $-40 \le b \le 40$, but show oscillatory behaviour for larger $b$, i.e. the convection dominated cases [3 & 5]. One of the treatments to the mentioned problem is to employ the so-called bubble function method, under the more general title of "stabilization techniques", in which the interpolation element includes some special element types in addition to the standard Lagrange elements. These elements are potentially useful for applications in fluid dynamics [1]. FEMLAB 3.1, employs these elements for application modes such as incompressible Navier-Stokes, Brinkman equations, non isothermal flow and many more, which is a novelty that could be extended to other eventualities. Setting $b=50$ and y=0.5, 0.7 the analytic solution looks like what follows:

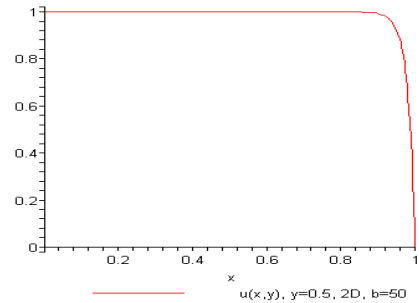

2.a Analytic result, layer y=0.5

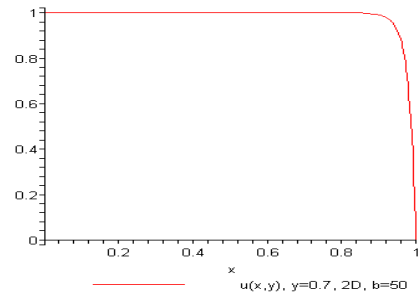

2.b Analytic result, layer y=0.7, x=0..1

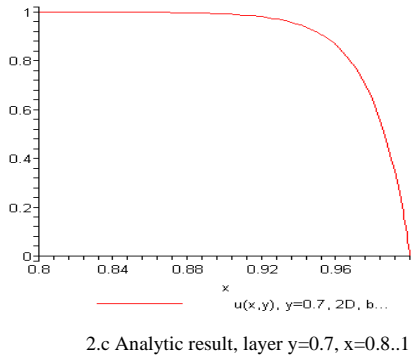

2.c Analytic result, layer y=0.7, x=0.8..1

**Figure 2. Two dimensional test problem plotted at layers y=0.5 (2.a), y=0.7 (2.b & 2.c)**

Solving the equation (3-3) using predefined quintic Lagrange element, FEMLAB returns within 4.438 seconds using a triangular mesh of 3976 elements with 50001 degrees of freedom. Figure 3 shows the plotted solution at different boundary layers:

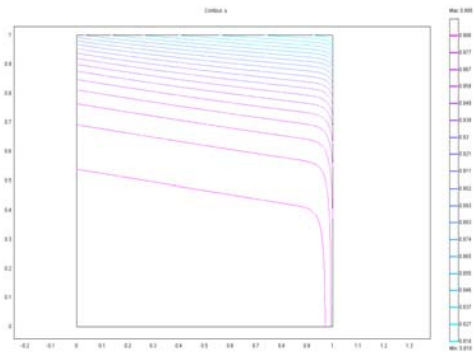

**Figure 3. FEMLAB 3.1 result, b=50, Various layers**

Equation (3-3) with large convection coefficients is known [5] to yield oscillatory results if treated by classical Galerkin finite element method. Different values of y should be interpreted on their own. FEMLAB results at this intermediate value of $b$ is, however, quite satisfactory and accurate.

## 4. Conclusions

A steady, scalar convection-diffusion model has been studied, in which the convection coefficient was dominant, although not very large compared to unity. Simulated results were highly reliable compared to the analytic solutions, processing and solution time was quite short and the software showed to be very easy to work with. The test problem was idealized in order to obtain analytic solutions, though, generalization to more complex geometries and problems is easily attainable. It was found that FEMLAB is a very successful modelling tool in terms of graphical features, coping with complex geometries, diversity of modules and models and specially being equipped with the stabilization techniques in many application modes.

**Acknowledgment**
The authors would like to thank Prof. V. Nassehi and Prof. R. J. Wakeman of Loughborough University, for their helps and supports.